\documentclass[reprint,amsmath,amssymb,superscriptaddress]{revtex4-1}

\usepackage{graphicx}
\usepackage{dcolumn}
\usepackage{bm}
\usepackage{color}

\begin{document}


\title{Interfacial giant tunnel magnetoresistance and bulk-induced large perpendicular magnetic anisotropy in (111)-oriented junctions with fcc ferromagnetic alloys: A first-principles study}


\author{Keisuke Masuda}
\email{MASUDA.Keisuke@nims.go.jp}
\affiliation{Research Center for Magnetic and Spintronic Materials, National Institute for Materials Science (NIMS), Tsukuba 305-0047, Japan}
\author{Hiroyoshi Itoh}
\affiliation{Department of Pure and Applied Physics, Kansai University, Suita 564-8680, Japan}
\affiliation{Center for Spintronics Research Network, Osaka University, Toyonaka 560-8531, Japan}
\author{Yoshiaki Sonobe}
\affiliation{Samsung R\,$\&$D Institute Japan, Yokohama 230-0027, Japan}
\author{Hiroaki Sukegawa}
\affiliation{Research Center for Magnetic and Spintronic Materials, National Institute for Materials Science (NIMS), Tsukuba 305-0047, Japan}

\author{Seiji Mitani}
\affiliation{Research Center for Magnetic and Spintronic Materials, National Institute for Materials Science (NIMS), Tsukuba 305-0047, Japan}
\affiliation{Center for Materials Research by Information Integration, National Institute for Materials Science (NIMS), Tsukuba 305-0047, Japan}
\affiliation{Graduate School of Pure and Applied Sciences, University of Tsukuba, Tsukuba 305-8577, Japan}

\author{Yoshio Miura}
\affiliation{Research Center for Magnetic and Spintronic Materials, National Institute for Materials Science (NIMS), Tsukuba 305-0047, Japan}
\affiliation{Center for Spintronics Research Network, Osaka University, Toyonaka 560-8531, Japan}
\affiliation{Center for Materials Research by Information Integration, National Institute for Materials Science (NIMS), Tsukuba 305-0047, Japan}


\date{\today}

\begin{abstract}
We study the tunnel magnetoresistance (TMR) effect and magnetocrystalline anisotropy in a series of magnetic tunnel junctions (MTJs) with $L1_1$-ordered fcc ferromagnetic alloys and MgO barrier along the [111] direction. Considering the (111)-oriented MTJs with different $L1_1$ alloys, we calculate their TMR ratios and magnetocrystalline anisotropies on the basis of the first-principles calculations. The analysis shows that the MTJs with Co-based alloys (CoNi, CoPt, and CoPd) have high TMR ratios over 2000\%. These MTJs have energetically favored Co-O interfaces where interfacial antibonding between Co $d$ and O $p$ states is formed around the Fermi level. We find that the resonant tunneling of the antibonding states, called the interface resonant tunneling, is the origin of the obtained high TMR ratios. Such a mechanism is similar to that found in our recent work on the simple Co/MgO/Co(111) MTJ [K. Masuda {\it et al}., Phys. Rev. B {\bf 101}, 144404 (2020)]. In contrast, different systems have different spin channels where the interface resonant tunneling occurs; for example, the tunneling mainly occurs in the majority-spin channel in the CoNi-based MTJ while it occurs in the minority-spin channel in the CoPt-based MTJ. This means that even though the mechanism is similar, different spin channels contribute dominantly to the high TMR ratio in different systems. Such a difference is attributed to the different exchange splittings in the particular Co $d$ states contributing to the tunneling though the antibonding with O $p$ states. Our calculation of the magnetocrystalline anisotropy shows that many $L1_1$ alloys have large perpendicular magnetic anisotropy (PMA). In particular, CoPt has the largest value of anisotropy energy $K_{\rm u} \approx 10\,{\rm MJ/m^3}$. We further conduct a perturbation analysis of the PMA with respect to the spin-orbit interaction and reveal that the large PMA in CoPt and CoNi mainly originates from spin-conserving perturbation processes around the Fermi level.
\end{abstract}


\maketitle

\section{\label{introduction} introduction}
Magnetic tunnel junctions (MTJs), in which an insulating tunnel barrier is sandwiched between ferromagnetic electrodes, have attracted considerable attention not only from the viewpoint of fundamental physics but also from their potential applications to various devices. In particular, for the application to nonvolatile magnetic random access memories (MRAMs), they need to have perpendicular magnetic anisotropy (PMA) as well as high tunnel magnetoresistance (TMR) ratios. The PMA is more beneficial than in-plane magnetic anisotropy for achieving high thermal stability when device sizes are scaled down in ultrahigh-density MRAMs \cite{2016Dieny-Wiley}. The PMA is also preferred for the different types of magnetization switching in MRAMs; the critical current for the switching in spin-transfer-torque MRAMs (STT-MRAMs) \cite{2016Dieny-Wiley} can be reduced and the write error rate in voltage-controlled MRAMs \cite{2016Shiota-APEX} can be decreased.

To obtain both large PMA and high TMR ratios in MTJs, two types of approaches have been employed. One approach is to utilize ferromagnets with strong bulk magnetocrystalline anisotropy as electrodes of MTJs. The ordered alloys, $L$1$_0$ FePt \cite{1995Klemmer-SMM,2002Okamoto-PRB}, $D$0$_{22}$ Mn$_3$Ga(Ge) \cite{2009Wu-APL,2011Mizukami-PRL,2012Kurt-APL,2013Mizukami-APEX}, and $L$1$_0$ MnGa \cite{2011Mizukami-PRL}, are ferromagnets with such strong magnetic anisotropy along the [001] direction, by which one can achieve large PMA in the (001)-oriented MTJs. However, unfortunately, these MTJs did not show high TMR ratios even if one of the ferromagnetic electrodes was replaced by CoFe(B) or Fe \cite{2008Yoshikawa-IEEE,2011Kubota-APEX,2012Ma-APL,2013Kubota-JPD,2016Lee-IEEE}. The other approach is to combine the interface-induced PMA and the established technology for high TMR ratios in Fe(Co)/MgO/Fe(Co)(001) MTJs \cite{2004Yuasa-NatMat,2004Parkin-NatMat}. Actually, experiments on CoFeB/MgO/CoFeB MTJs \cite{2010Ikeda-NatMat} demonstrated relatively large interfacial PMA ($\sim 1.3\,{\rm mJ/m^2}$) and high TMR ratios ($>120\%$ at room temperature). However, such an interfacial PMA is sensitive to the interfacial oxidation condition \cite{2011Yang-PRB,2013Hallal-PRB} and the thickness of the ferromagnetic layers \cite{2010Ikeda-NatMat}. Thus, large PMA due to bulk magnetocrystalline anisotropy is attractive for storage layers of MRAMs. It should also be remarked that large bulk PMA is beneficial for the pinned layers in the synthetic antiferromagnetic structures in MRAM cells \cite{2017Yakushiji-APL}. In this study, we theoretically demonstrate such large bulk-induced PMA and high TMR ratios in unconventional MTJs and discuss their physical underlying mechanisms.

\begin{figure}
\includegraphics[width=8.7cm]{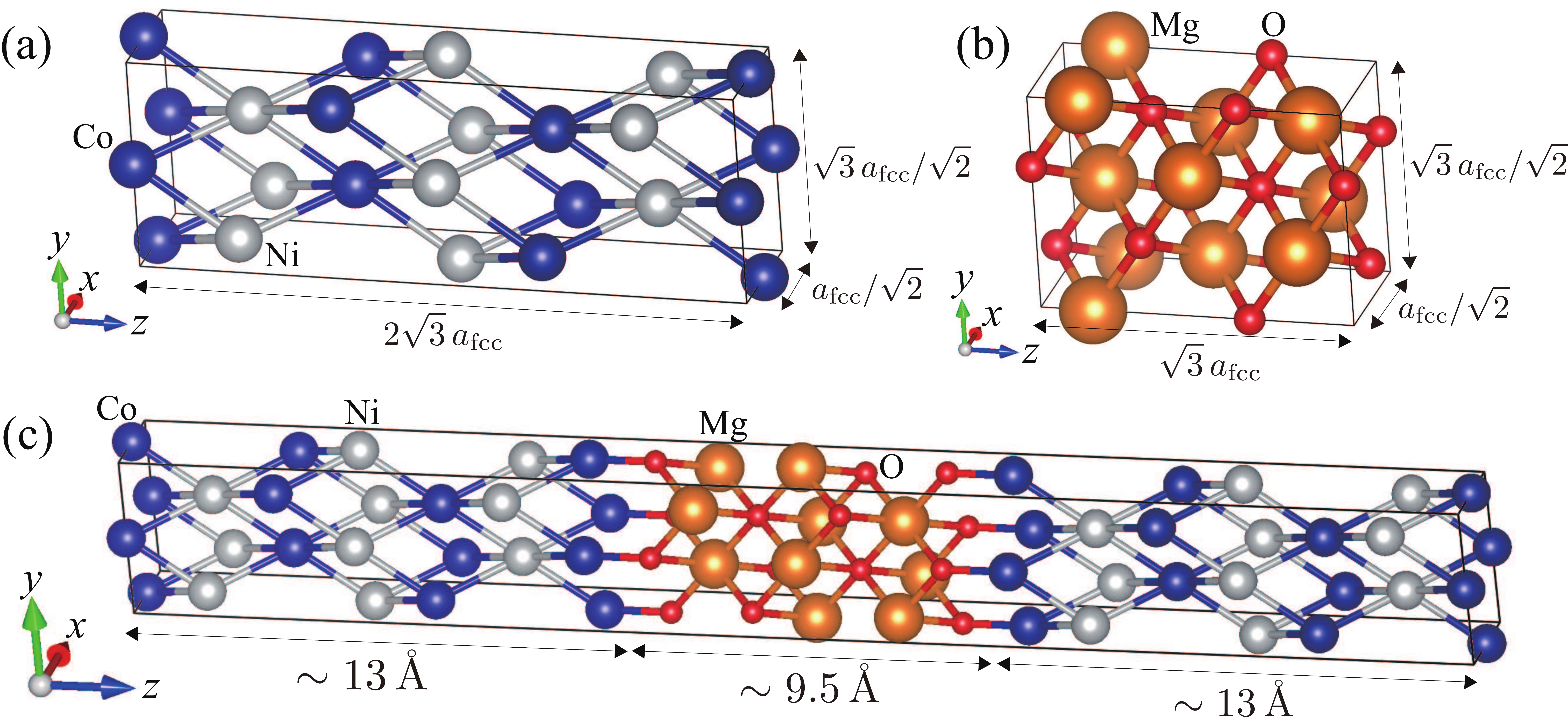}
\caption{\label{crystal-structures} The unit cells of (a) $L1_1$ CoNi and (b) MgO, where the $z$ axes are set to their [111] directions of the original fcc cells. (c) The supercell of ${\rm CoNi}\, (7\, {\rm ML})/{\rm MgO}\, (7\, {\rm ML})/{\rm CoNi}\, (7\, {\rm ML})(111)$.}
\end{figure}

Let us here introduce unconventional (111)-oriented MTJs, where fcc ferromagnetic electrodes and the fcc tunnel barrier are stacked along their [111] directions [Fig. \ref{crystal-structures}(c)]. It is natural to consider such (111)-oriented MTJs for fcc materials, since the (111) plane is the close-packed plane of the fcc lattice and has the lowest surface energy \cite{1988Ting-SurfSci}. However, most previous studies addressed (001)-oriented MTJs with bcc materials because of the initial success in Fe/MgO/Fe(001) \cite{2004Yuasa-NatMat,2004Parkin-NatMat,2001Butler-PRB,2001Mathon-PRB}. Recently, three of the present authors theoretically investigated the TMR effect in two simple (111)-oriented MTJs, Co/MgO/Co(111) and Ni/MgO/Ni(111), and obtained a high TMR ratio ($\sim 2100\%$) in the Co-based MTJ \cite{2020Masuda-PRB}. This result motivates us to study other (111)-oriented MTJs for obtaining high TMR ratios.

Another important merit of (111)-oriented MTJs is that several magnetic superlattices and $L1_1$ alloys can be used as ferromagnetic electrodes for large PMA \cite{2017Seki-JPSJ,1992Johnson-PRL,2008Sato-JAP,2010Yakushiji-APL,2009Mizunuma-APL}. For example, Seki {\it et al.} \cite{2017Seki-JPSJ} recently observed large PMA with uniaxial magnetic anisotropic energy ($K_{\rm u}$) of $\sim 0.5\, {\rm MJ}/{\rm m}^3$ in epitaxial Co/Ni(111) multilayers, consistent with previous experiments \cite{1992Johnson-PRL}. In another experimental study \cite{2008Sato-JAP}, Sato {\it et al.} grew $L$1$_1$ CoPt films on an MgO(111) substrate and showed large PMA ($K_{\rm u}\sim 3.7\, {\rm MJ}/{\rm m}^3$). Furthermore, Yakushiji {\it et al.} \cite{2010Yakushiji-APL} obtained PMA ($K_{\rm u}\sim 0.5\, {\rm MJ}/{\rm m}^3$) in Co/Pt(111) and Co/Pd(111) multilayers that have similar structures as $L1_1$ films. All these studies indicate the potential of (111)-oriented MTJs with $L1_1$ alloys for large PMA; however, such MTJs have not been investigated both theoretically and experimentally in previous studies.

In this work, we present a systematic theoretical study of the TMR effect and magnetocrystalline anisotropy in (111)-oriented MTJs with $L1_1$ alloys. We consider various possible MTJs consisting of $L1_1$ alloys and the MgO tunnel barrier and calculate their TMR ratios and magnetocrystalline anisotropies by means of the first-principles calculations. It is shown that the MTJs with Co-based alloys (CoNi, CoPt, and CoPd) have high TMR ratios over 2000\%. The detailed analysis of the electronic structures and conductances clarifies that all the obtained high TMR ratios originate from the resonant tunneling of the interfacial $d$-$p$ antibonding states called the interface resonant tunneling \cite{2020Masuda-PRB}, which is clearly different from the conventional mechanism of the high TMR ratio in Fe/MgO/Fe(001) \cite{2001Butler-PRB,2001Mathon-PRB}. The interface resonant tunneling mainly occurs in the majority- and minority-spin channels in the CoNi- and CoPt-based MTJs, respectively. Namely, the high TMR ratios in different systems come from the tunneling in different spin channels. In the calculation of the magnetocrystalline anisotropy, we obtain large PMA in many $L1_1$ alloys. Among them, CoPt has the largest $K_{\rm u}$ of $\approx 10\,{\rm MJ/m^3}$. A second-order perturbation analysis of the PMA with respect to the spin-orbit interaction (SOI) clarifies that the large PMA in CoPt and CoNi originates from the spin-conserving perturbation processes around the Fermi level.

\section{\label{modelmethod} model and method}
\subsection{Structure optimization}
\begin{table}
\caption{\label{tab1}
The optimized value of $a_{\rm fcc}$ in each $L1_1$ alloy and the calculated TMR ratio in the corresponding (111)-oriented MTJ. The anisotropy energy $K_{\rm u}$ calculated in each $L1_1$ alloy is also shown. We calculated TMR ratios using supercells with 7 ML of MgO. Only the TMR ratio in the bottom row was calculated for the thicker barrier (MgO 13 ML). The values of $K_{\rm u}$ are given in units of MJ/m$^3$ (meV/cell). These $K_{\rm u}$ values were calculated using the unit cell of each $L1_1$ alloy [Fig. \ref{crystal-structures}(a)] including 12 atoms.
}
\begin{ruledtabular}
\renewcommand{\arraystretch}{1.2}
\begin{tabular}{cccc}
&
\textrm{$a_{\rm fcc}$\,(\AA)}&
\textrm{TMR ratio\,(\%)}&
\textrm{$K_{\rm u}$}\\
\colrule
FePt & 3.83 & 716  &  4.95 (5.21)  \\
CoPt & 3.79 & 2534 &  9.86 (10.04) \\
NiPt & 3.78 & 650  & --1.04 (--1.05) \\
FePd & 3.81 & 46   &  0.73 (0.76)  \\
CoPd & 3.76 & 2172 &  1.88 (1.87)  \\
NiPd & 3.76 & 585  &  0.45 (0.45)  \\
FeNi & 3.56 & 484  &  0.67 (0.56)  \\
CoNi & 3.51 & 3210 &  1.10 (0.89)  \\
CoNi & 3.51 & 2361 $({\rm MgO}\,13\,{\rm ML})$ &  1.10 (0.89)  \\
\end{tabular}
\end{ruledtabular}
\end{table}
Since the $L1_1$ phase can exist only in multilayer films owing to its metastable nature, it is hard to obtain the experimental lattice constants of the $L1_1$ alloys. This forces us to conduct the structure optimization to theoretically determine the optimal lattice constants. In the present study, we considered eight different $L1_1$ alloys (Table \ref{tab1}) and prepared their unit cells with the $z$ axis along the [111] direction of the original fcc cell [Fig. \ref{crystal-structures}(a)]. We optimized the value of $a_{\rm fcc}$ in each $L1_1$ alloy by means of the density-functional theory (DFT) implemented in the Vienna {\it ab initio} simulation program ({\scriptsize VASP}) \cite{1996Kresse-PRB}. Here, we adopted the generalized gradient approximation (GGA) \cite{1996Perdew-PRL} for the exchange-correlation energy and used the projected augmented wave (PAW) pseudopotential \cite{1994Bloechl-PRB,1999Kresse-PRB} to treat the effect of core electrons properly. A cutoff energy of 337\,eV was employed and the Brillouin-zone integration was performed with $23\times13\times5$ {\bf k} points. The convergence criteria for energy and force were set to $10^{-5}\,{\rm eV}$ and $10^{-4}\,{\rm eV/\AA}$, respectively. The obtained values of $a_{\rm fcc}$ are shown in Table \ref{tab1}.

By combining the unit cell of each $L1_1$ alloy [Fig. \ref{crystal-structures}(a)] and that of the (111)-oriented MgO [Fig. \ref{crystal-structures}(b)], we built the supercell of the corresponding (111)-oriented MTJ [Fig. \ref{crystal-structures}(c)]. The $x$- and $y$-axis lengths of the supercell were fixed to $a_{\rm fcc}/\sqrt{2}$ and $\sqrt{3}\,a_{\rm fcc}/\sqrt{2}$ in each supercell where the optimized $a_{\rm fcc}$ of each alloy was used. The atomic positions along the $z$ direction in the supercells were relaxed using the DFT with the aid of the {\scriptsize VASP} code. In these calculations for supercells, $23\times13\times1$ {\bf k} points were used, and the other calculation conditions were the same as the structure optimizations of the $L1_1$ alloys. More technical details of structure optimizations of supercells are given in our previous work \cite{2017Masuda-PRB_bias}. In each supercell, we compared energies for all interfacial atomic configurations and determined the energetically favored configuration. For example, in CoNi/MgO/CoNi(111), there are four atomic configurations at the interface: Co-O, Ni-O, Co-Mg, and Ni-Mg. By comparing formation energies for these cases, we found that the Co-O interface has the lowest energy. In Table \ref{tab1}, each $L1_1$-ordered alloy is denoted as $XY$ ($X={\rm Co}$ and $Y={\rm Ni}$ for CoNi). We confirmed that the $X$-O interface was energetically favored in each supercell. Such supercells with energetically favored interfaces were used in the transport calculation explained below.

\subsection{Calculation method of TMR ratios}
\begin{figure}
\includegraphics[width=8.2cm]{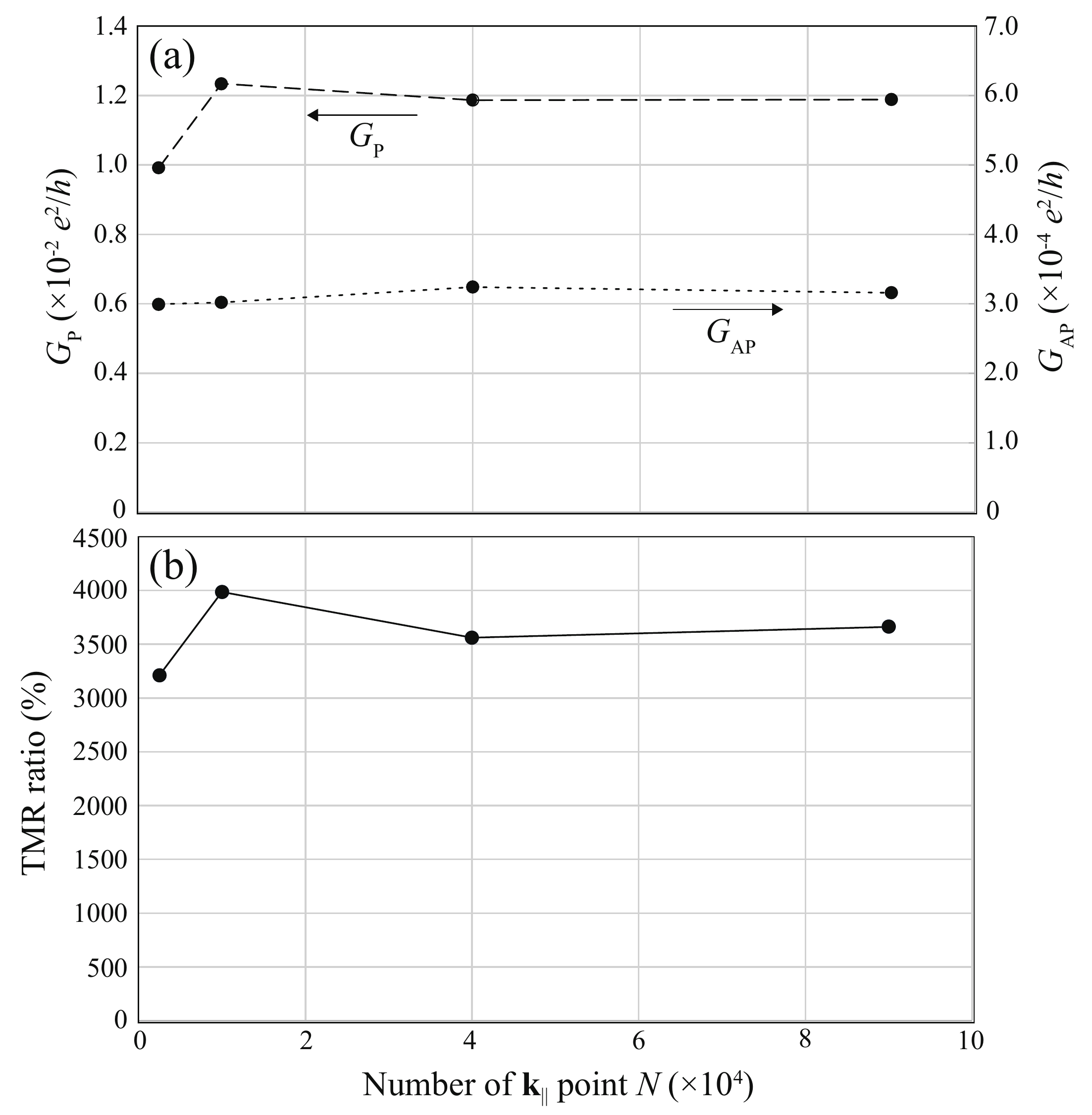}
\caption{\label{kptnumber_dep} The ${\bf k}_{\parallel}$-point number $N$ dependencies of (a) $G_{\rm P}$ (dashed line) and $G_{\rm AP}$ (dotted line) and (b) the TMR ratio in ${\rm CoNi}/{\rm MgO}\,(7\,{\rm ML})/{\rm CoNi}(111)$.}
\end{figure}
The TMR ratio of each (111)-oriented MTJ was calculated using the DFT and Landauer formula with the help of the {\scriptsize PWCOND} code \cite{2004Smogunov-PRB} in the {\scriptsize QUANTUM ESPRESSO} package \cite{Baroni}. We first constructed the quantum open system by attaching the left and right semi-infinite electrodes of each $L$1$_1$ alloy to the supercell. Here, the supercell is composed of seven monolayers (ML) of MgO sandwiched between 7 ML of $L1_1$ alloys $L1_1\, (7\, {\rm ML})/{\rm MgO}\, (7\, {\rm ML})/L1_1\, (7\, {\rm ML})$ [Fig. \ref{crystal-structures}(c)] for the parallel configuration of the magnetization, whose atomic positions are optimized following the procedures mentioned in Sec. \ref{modelmethod} A. The thickness of the MgO barrier layers is about $9.5\, {\rm \AA}$. In the case of the antiparallel magnetization, we need to use a supercell that is twice as long as that for the parallel magnetization to satisfy the translational invariance in the magnetization state along the stacking direction; this is made by connecting two $L1_1\, (7\, {\rm ML})/{\rm MgO}\, (7\, {\rm ML})/L1_1\, (7\, {\rm ML})$ cells inverting one of them. The application of the DFT to the quantum open system provided the self-consistent potential, which was used to derive the scattering equation mentioned below. In the DFT calculation, the exchange-correlation energy was treated within the GGA, and the ultrasoft pseudopotentials were used. The cutoff energies were set to 45 and 450\,Ry for the wave function and the charge density, respectively. The number of ${\bf k}$ points was taken to be 23$\times$13$\times$1 and the convergence criterion was set to $10^{-6}\,{\rm Ry}$. Since our system has translational symmetry in the $xy$-plane, the scattering states can be classified by an in-plane wave vector ${\bf k}_{\parallel}=(k_x,k_y)$. For each ${\bf k}_{\parallel}$ and spin index, we solved the scattering equation derived under the condition that the wave function and its derivative of the supercell are connected to those of the electrodes \cite{1999Choi-PRB,2004Smogunov-PRB}. These calculations gave the ${\bf k}_{\parallel}$-resolved transmittances from which the ${\bf k}_{\parallel}$-resolved conductances were obtained through the Landauer formula: $G_{{\rm P},\uparrow}({\bf k}_{\parallel})$, $G_{{\rm P},\downarrow}({\bf k}_{\parallel})$, $G_{{\rm AP},\uparrow}({\bf k}_{\parallel})$, and $G_{{\rm AP},\downarrow}({\bf k}_{\parallel})$. Here, ${\rm P}$ (${\rm AP}$) refers to the parallel (antiparallel) magnetization state of the electrodes and $\uparrow$ ($\downarrow$) indicates the majority-spin (minority-spin) channel. Note that in the antiparallel state, we defined the index $\sigma$ for $G_{{\rm AP},\sigma}({\bf k}_{\parallel})$ as the channel for the left electrode; namely, $G_{{\rm AP},\uparrow}({\bf k}_{\parallel})$ [$G_{{\rm AP},\downarrow}({\bf k}_{\parallel})$] corresponds to the electron tunneling from the majority-spin (minority-spin) channel in the left electrode to the minority-spin (majority-spin) channel in the right electrode. We averaged each conductance over ${\bf k}_{\parallel}$ as, e.g., $G_{{\rm P},\uparrow}=\sum_{{\bf k}_\parallel}G_{{\rm P},\uparrow}({\bf k}_\parallel)/N$, where $N$ is the sampling number of ${\bf k}_{\parallel}$ points. For each MTJ, we calculated the TMR ratio following its optimistic definition:
\begin{equation}
{\rm TMR\,\, ratio}\,\, (\%) =100 \times (G_{\rm P}-G_{\rm AP})/G_{\rm AP},\label{eq:TMR}
\end{equation}
where $G_{\rm P(AP)}=G_{{\rm P(AP)},\uparrow}+G_{{\rm P(AP)},\downarrow}$. In the present work, we neglect the SOI in the calculation of TMR ratios. This is because, as will be discussed in Sec. \ref{resultsdiscussion} B, it is expected that the SOI does not affect TMR ratios significantly at least for the present MTJs with Co-O interfaces that exhibit high TMR ratios.

We carefully considered ${\bf k}_{\parallel}$-point number $N$ dependencies of $G_{\rm P}$, $G_{\rm AP}$, and the TMR ratio. Figure \ref{kptnumber_dep} shows these quantities in CoNi/MgO/CoNi(111) as a function of $N$, from which we found that $N \geq 40\,000$ is required for the convergence of these quantities. In the following part of the manuscript, we show the results calculated with $N=40\,000$.

\subsection{Estimation of magnetocrystalline anisotropy}
We calculated the uniaxial magnetic anisotropy energy $K_{\rm u}$ of each $L$1$_1$ alloy on the basis of the DFT calculation including the SOI. We adopted the expression by the well-known force theorem \cite{1990Daalderop-PRB,1985Weinert-PRB}:
\begin{equation}
K_{\rm u}=(E_{\parallel}-E_{\perp})/V,\label{eq:Ku}
\end{equation}
where $E_{\parallel}$ ($E_{\perp}$) is the sum of the eigenvalues only over occupied states of the unit cell [Fig. \ref{crystal-structures}(a)] with the magnetization along the $x$ ($z$) direction, and $V$ is the volume of the unit cell. Here, we used the optimized lattice constant mentioned above for each $L$1$_1$ alloy. From the definition in Eq. (\ref{eq:Ku}), a positive (negative) $K_{\rm u}$ indicates a tendency toward PMA (in-plane magnetic anisotropy). The {\scriptsize VASP} code was used for the DFT calculation including the SOI, where we adopted the GGA for the exchange-correlation energy, the PAW pseudopotential, and a cutoff energy of 337\,eV. Since the energy scale of $K_{\rm u}$ is much smaller than that of the total energy of the system, the large number of {\bf k} points are required to estimate $K_{\rm u}$ accurately. We thus used $51\times27\times11$ {\bf k} points after confirming the convergence of $K_{\rm u}$ with respect to the number of {\bf k} points.

In addition to these calculations, we also conducted a second-order perturbation analysis of the magnetocrystalline anisotropy \cite{2013Miura-JPCM} to understand the origin of the PMA. By treating the SOI as a perturbation term, the second-order perturbation energy is given by
\begin{eqnarray}
  E^{(2)} &=& \sum_{\bm k} \sum^{\rm unocc}_{n'\sigma'} \sum^{\rm occ}_{n\sigma} \frac{|\langle {\bm k}n'\sigma' | H_{\rm SOI} | {\bm k}n\sigma \rangle|^{2}}{\epsilon^{(0)}_{{\bm k}n\sigma}-\epsilon^{(0)}_{{\bm k}n'\sigma'}}, \label{eq:2ndPT}\\
  H_{\rm SOI} &=& \sum_{i} \xi_{i}\, {\bm L}_{i} \cdot {\bm S}_{i},
\end{eqnarray}
where $\epsilon^{(0)}_{{\bm k}n\sigma}$ is the energy of an unperturbed state $| {\bm k}n\sigma \rangle$ with wave vector ${\bm k}$, band index $n$, and spin $\sigma$. The index ``occ'' (``unocc'') on the summation in Eq. (\ref{eq:2ndPT}) means that the sum is over occupied (unoccupied) states of all atoms in the unit cell. In the Hamiltonian for the SOI $H_{\rm SOI}$, $\xi_{i}$ is its coupling constant at an atomic site $i$, and ${\bm L}_i$ (${\bm S}_i$) is the single-electron angular (spin) momentum operator. Wave functions and eigenenergies obtained in our DFT calculations were used as unperturbed states and energies in Eq. (\ref{eq:2ndPT}). The magnetocrystalline anisotropy energy within the second-order perturbation was calculated as $E^{(2)}_{\rm MCA}=E^{(2)}_{\parallel}-E^{(2)}_{\perp}$similar to Eq. (\ref{eq:Ku}), where $E^{(2)}_{\parallel}$ ($E^{(2)}_{\perp}$) is the energy calculated by Eq. (\ref{eq:2ndPT}) for the magnetization along the $x$ ($z$) direction of the unit cell. We can decompose $E^{(2)}_{\rm MCA}$ into four types of terms coming from different perturbation processes at each atomic site:
\begin{eqnarray}
E^{(2)}_{\rm MCA}&=&\sum_{i} E^{i}_{\rm MCA},\\
E^{i}_{\rm MCA}&=&\Delta E^{i}_{\uparrow \Rightarrow \uparrow} + \Delta E^{i}_{\downarrow \Rightarrow \downarrow} + \Delta E^{i}_{\uparrow \Rightarrow \downarrow} + \Delta E^{i}_{\downarrow \Rightarrow \uparrow}. \label{eq:E2MCA_decomp}
\end{eqnarray}
Here, $E^{i}_{\rm MCA}$ is the magnetocrystalline anisotropy energy at each atomic site $i$. The term $\Delta E^{i}_{\uparrow \Rightarrow \uparrow}$ ($\Delta E^{i}_{\downarrow \Rightarrow \downarrow}$) is the contribution from spin-conserving perturbation processes in the majority-spin (minority-spin) channel. The last two terms are the contributions from spin-flip perturbation processes: $\Delta E^{i}_{\uparrow \Rightarrow \downarrow}$ ($\Delta E^{i}_{\downarrow \Rightarrow \uparrow}$) comes from electron transition processes from majority- to minority-spin (minority- to majority-spin) channel. This decomposition provides us with information on the origin of the PMA.

\section{\label{resultsdiscussion} results and discussion}
\subsection{High TMR ratios and their possible origin}
\begin{figure}
\includegraphics[width=7.6cm]{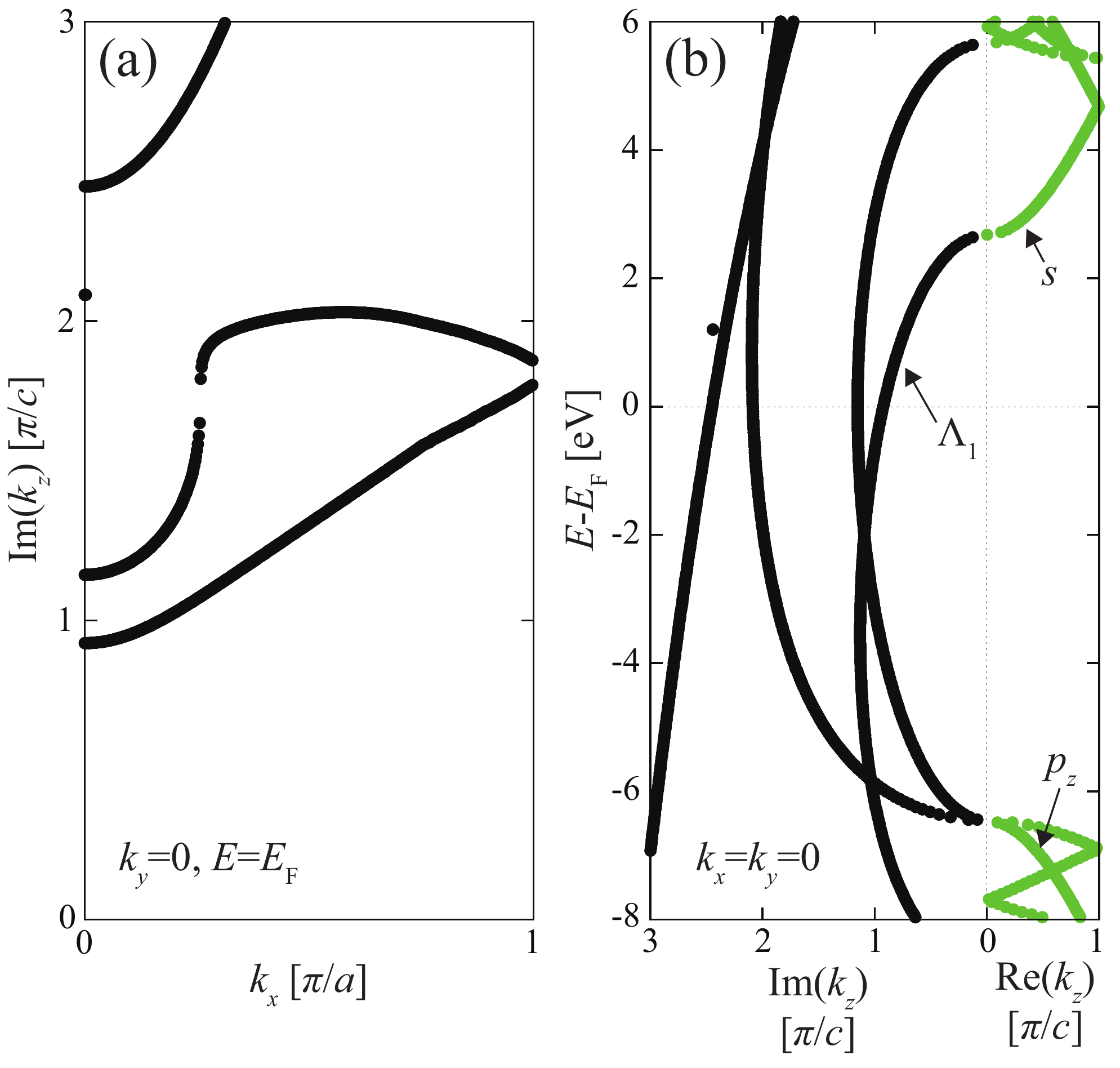}
\caption{\label{cband-mixed} Imaginary and real parts of $k_z$ calculated for the MgO unit cell [Fig. \ref{crystal-structures}(b)]. (a) Imaginary part of $k_z$ as a function of real $k_x$ ($k_y=0$) at the Fermi level $E_{\rm F}$. (b) Imaginary and real parts of $k_z$ around $E_{\rm F}$ at $k_x=k_y=0$.}
\end{figure}
Table \ref{tab1} shows the obtained TMR ratios in the (111)-oriented MTJs. The MTJs, including the Co-based alloys, have high TMR ratios over 2000\%. In contrast, the Fe- and Ni-based alloys give much lower TMR ratios ($<1000\%$).

To understand the origin of the high TMR ratios, the bulk band structures of the electrodes and the barrier were first analyzed because the high TMR ratio in the well-known Fe/MgO/Fe(001) MTJ \cite{2004Yuasa-NatMat,2004Parkin-NatMat} was explained by the bulk band structures of Fe and MgO on the basis of the coherent tunneling mechanism \cite{2001Butler-PRB,2001Mathon-PRB}. If a similar mechanism holds for the present MTJs, the bulk band structures along the $\Lambda$ line in the Brillouin zone corresponding to the [111] direction should explain the high TMR ratios.

Figure \ref{cband-mixed}(a) shows the imaginary part of $k_z$, referred to as the complex band, of the (111)-oriented MgO [Fig. \ref{crystal-structures}(b)] as a function of $k_x$. The smallest value of ${\rm Im}(k_z)$ is located at $(k_x,k_y)=(0,0)=\Gamma$. This means that the $\Lambda$ states, i.e., the wave function in the $\Lambda$ line $(0,0,k_z)$, has the slowest decay and can provide the dominant contribution to the electron transport. In Fig. \ref{cband-mixed}(b), we show the complex and real bands at the $\Lambda$ line. We find that the smallest ${\rm Im}(k_z)$ at $E_{\rm F}$ comes from the $\Lambda_1$ state consisting of $s$ and $p_z$ orbitals. Therefore, the $\Lambda_1$ state decays most slowly in the barrier and the selective transport of this state can occur.

\begin{figure}
\includegraphics[width=7.6cm]{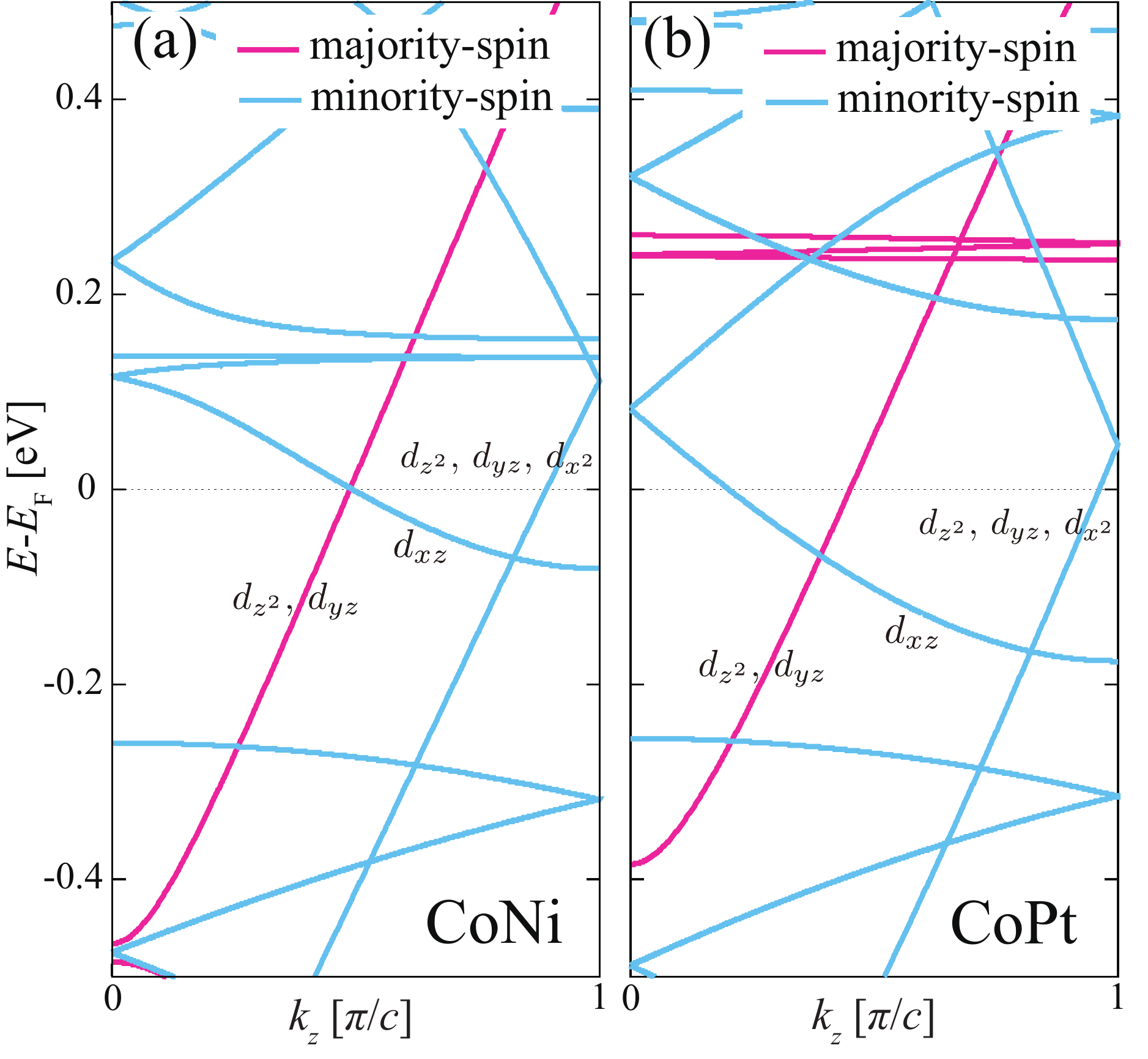}
\caption{\label{rband-mixed} Band structures along the $\Lambda$ line of (a) $L1_1$ CoNi and (b) $L1_1$ CoPt. In both panels, atomic orbitals contributing dominantly to each band around $E_{\rm F}$ are indicated, where $d_{3z^2-r^2}$ and $d_{x^2-y^2}$ are abbreviated as $d_{z^2}$ and $d_{x^2}$, respectively.}
\end{figure}
To study whether the $L1_1$ alloys have half-metallicity in the $\Lambda_1$ state, bulk band structures of CoNi and CoPt, which provide the two highest TMR ratios, were analyzed. As shown in Figs. \ref{rband-mixed}(a) and \ref{rband-mixed}(b), both majority- and minority-spin bands from the $d_{3z^2-r^2}$ state (belonging to the $\Lambda_1$ state) cross the Fermi level in both alloys; namely, these alloys do not have half-metallicity in the $\Lambda_1$ state, which is in sharp contrast to the half-metallicity in the $\Delta_1$ state of Fe in Fe/MgO/Fe(001) \cite{2001Butler-PRB,2001Mathon-PRB}. All these results indicate that we cannot explain the present high TMR ratios from the bulk band structures based on the coherent tunneling mechanism as in Fe/MgO/Fe(001).

\begin{figure}
\includegraphics[width=8.5cm]{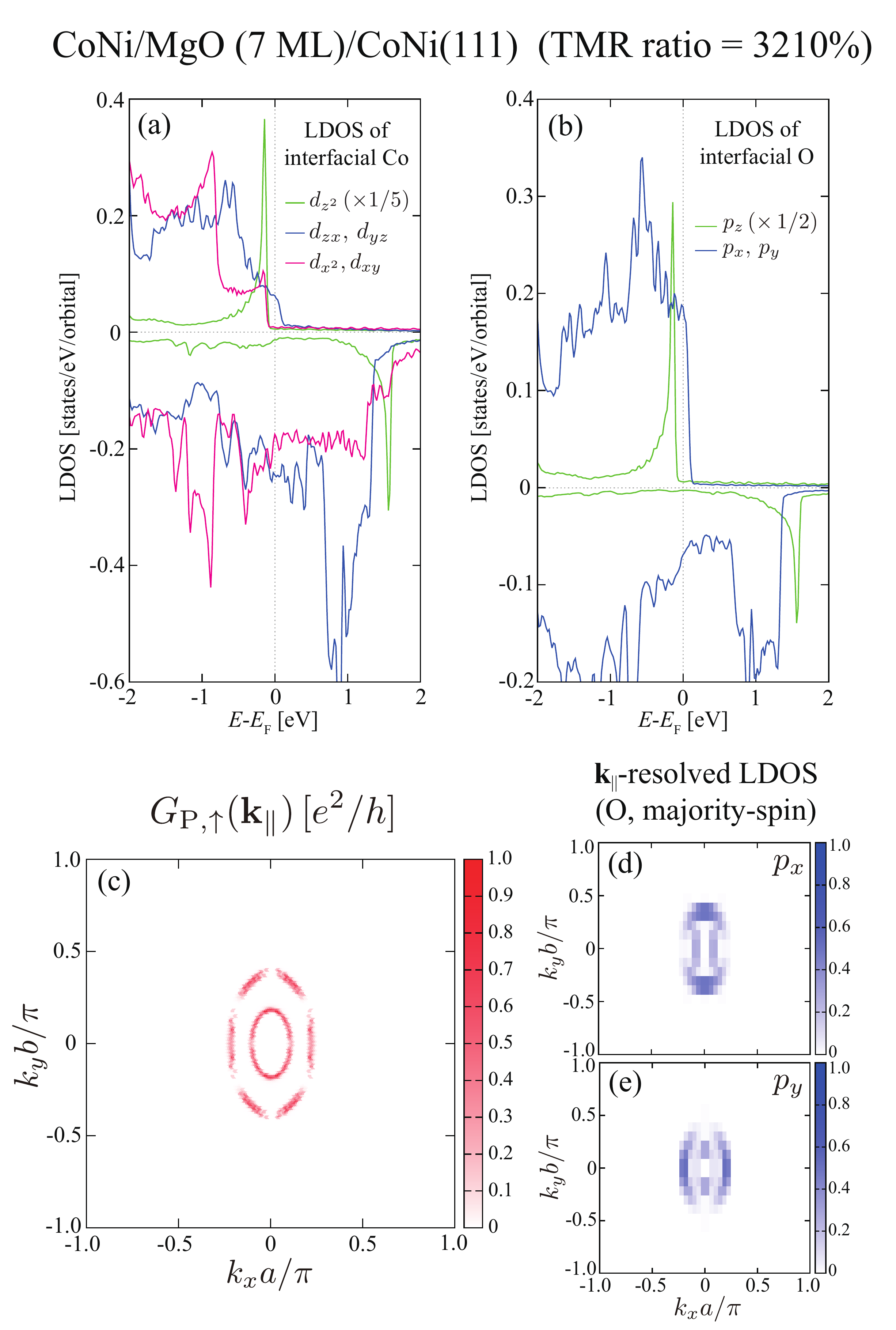}
\caption{\label{CoNiMgO} The electronic structures and transport properties of ${\rm CoNi}/{\rm MgO}\,(7\,{\rm ML})/{\rm CoNi}(111)$. (a,\,b) Projected LDOSs at interfacial Co and O atoms, where $d_{3z^2-r^2}$ and $d_{x^2-y^2}$ are abbreviated as $d_{z^2}$ and $d_{x^2}$, respectively. (c) The ${\bf k}_{\parallel}$ dependence of the majority-spin conductance in the parallel configuration of magnetizations. (d,\,e) The ${\bf k}_{\parallel}$-resolved LDOSs at $E=E_{\rm F}$ in the majority-spin channel projected onto the $p_x$ and $p_y$ states of interfacial O atoms.}
\end{figure}
\begin{table}
\caption{\label{tab2}
The conductances $G_{\rm P}=G_{\rm P,\uparrow}+G_{\rm P,\downarrow}$ and $G_{\rm AP}=G_{\rm AP,\uparrow}+G_{\rm AP,\downarrow}$ obtained using ${\rm CoNi}/{\rm MgO}\,(n\,{\rm ML})/{\rm CoNi}$ ($n=7,\,13$) supercells.
}
\begin{ruledtabular}
\renewcommand{\arraystretch}{1.4}
\begin{tabular}{ccc}
MgO thickness & 7\,ML\, (9.5\,\AA) & 13\,ML\, (19\,\AA) \\
\colrule
$G_{\rm P}\,(e^2/h)$ & $1.19 \times 10^{-2}$ & $2.64 \times 10^{-6}$ \\
$G_{\rm AP}\,(e^2/h)$ & $3.24 \times 10^{-4}$ & $1.07 \times 10^{-7}$ \\
\textrm{TMR ratio\,(\%)} & 3561 & 2361 \\
\end{tabular}
\end{ruledtabular}
\end{table}
\begin{figure}
\includegraphics[width=8.5cm]{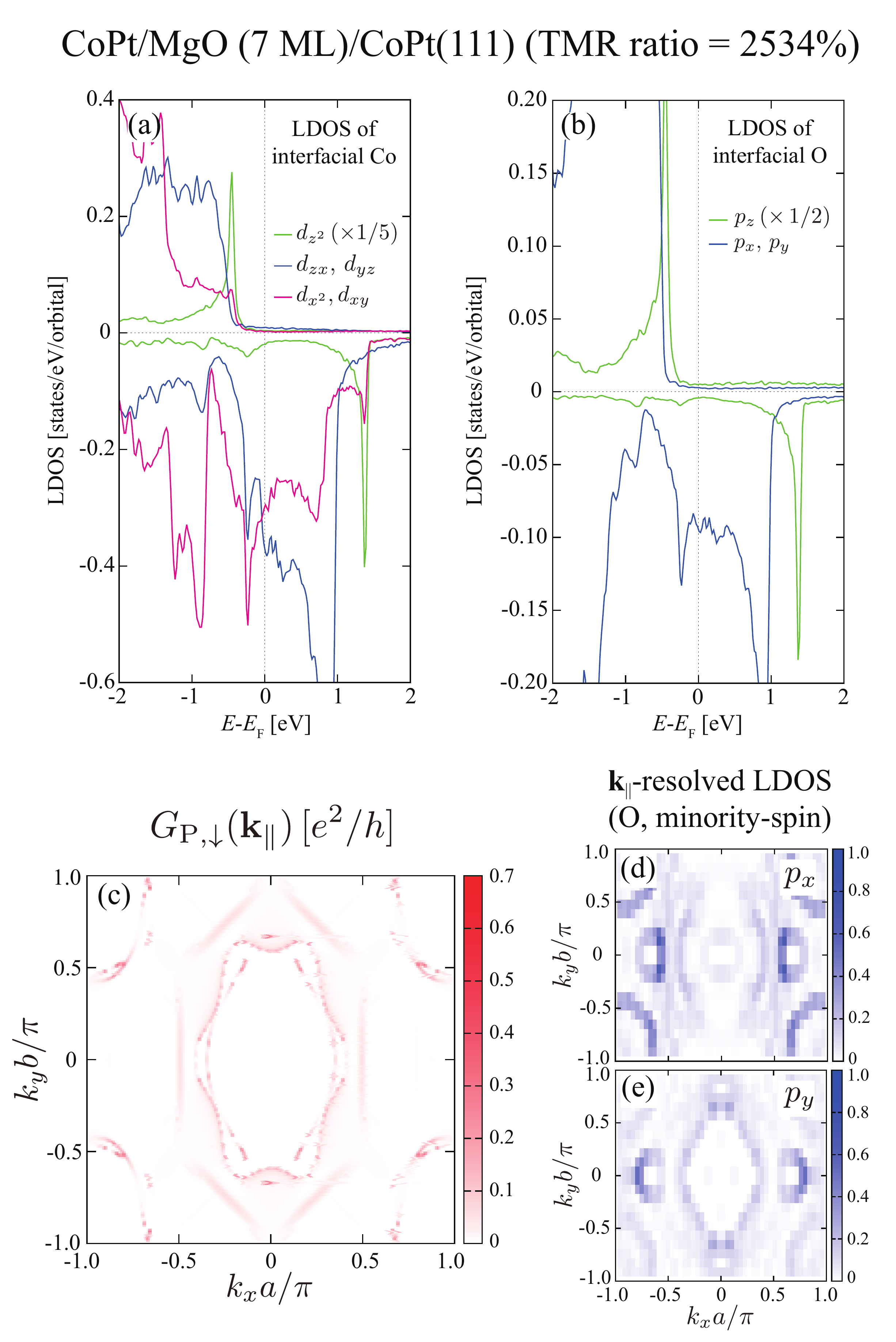}
\caption{\label{CoPtMgO} The same as Fig \ref{CoNiMgO}, but for ${\rm CoPt}/{\rm MgO}\,(7\,{\rm ML})/{\rm CoPt}(111)$. Note that the conductance and LDOSs in the minority-spin channel are shown in panels (c) to (e).}
\end{figure}
Another possible way to understand the present high TMR ratios is to focus on interfacial effects. In our previous study \cite{2020Masuda-PRB}, we clarified that the interface resonant tunneling provides a high TMR ratio in a simple (111)-oriented MTJ, Co/MgO/Co(111). To examine a similar possibility, we calculated the local density of states (LDOSs) at interfacial Co and O atoms of CoNi/MgO/CoNi(111) shown in Figs. \ref{CoNiMgO}(a) and \ref{CoNiMgO}(b). We can find a clear similarity in the energy dependence of the LDOS between the Co $d_{zx}$ ($d_{yz}$) and O $p_x$ ($p_y$) states in the majority-spin channel due to the formation of the interfacial antibonding between these states. At the Fermi level, such O $p_x$ and $p_y$ states have large LDOSs and can provide interfacial resonant tunneling between the left and right interfaces. Figure \ref{CoNiMgO}(c) shows the ${\bf k}_\parallel$-resolved conductance $G_{{\rm P},\uparrow}({\bf k}_\parallel)$, which contributes dominantly to the high TMR ratio. The conductance has only a small value at ${\bf k}_{\parallel}=\Gamma$, and their large values distribute around the $\Gamma$ point, which is a characteristic in the conductance originating from interfacial effects. We also analyzed the ${\bf k}_\parallel$-resolved LDOSs of the interfacial O $p_x$ and $p_y$ majority-spin states as shown in Figs. \ref{CoNiMgO}(d) and \ref{CoNiMgO}(e). The distribution of ${\bf k}_\parallel$ points with large LDOS is similar to that with large conductance in Fig. \ref{CoNiMgO}(c), indicating that the interfacial O $p_x$ and $p_y$ states play the dominant role in the high TMR ratio through the interfacial resonant tunneling.

As shown in the bottom of Table \ref{tab1}, we also calculated the TMR ratio in the thicker MgO case ($\sim 19\,{\rm \AA}$) using a supercell ${\rm CoNi}/{\rm MgO}\,(13\,{\rm ML})/{\rm CoNi}$. A high TMR ratio over $2000\%$ was obtained for this thicker barrier, although the value is lower than the thinner barrier case. When the MgO barrier becomes thicker, the interfacial resonant tunneling is weakened depending on the value of the smallest ${\rm Im}(k_z)$ [Fig. \ref{cband-mixed}(a)] at the ${\bf k}_{\parallel}$ points with interfacial resonance states. Actually, as shown in Table \ref{tab2}, the total conductance $G_{\rm P}$ in the parallel magnetization state decreases to $2.64 \times 10^{-6}\,e^2/h$ by increasing the number of MgO layers. However, since the conductance $G_{\rm AP}$ in the antiparallel magnetization state also largely diminishes, the TMR ratio has a high value ($>2000\%$) as mentioned above. Although not shown here, we confirmed that the ${\bf k}_\parallel$ dependence of the conductance $G_{{\rm P},\uparrow}({\bf k}_\parallel)$ in the thicker MgO case is almost the same as that in the thinner MgO case [Fig. \ref{CoNiMgO}(c)], indicating that the interfacial resonant tunneling is still active for the thicker MgO. The decay of the parallel conductance $G_{\rm P}$ with increasing the MgO thickness (Table \ref{tab2}) can be roughly estimated from the complex band shown in Fig. \ref{cband-mixed}. When we use complex wave vector $\kappa={\rm Im}(k_z)=0.94\,\pi/c$ for simplicity, the decay factor for the conductance is calculated as $\exp(-2\kappa d) \approx 9.81 \times 10^{-5}$, where we used $d=9.5\,{\rm \AA}$ as the increment in the MgO thickness ($7 \rightarrow 13\,{\rm ML}$) and $c=\sqrt{3} \times 3.51\,{\rm \AA}$ as the $c$-axis length of the MgO cell [Fig. \ref{crystal-structures}(b)]. Using this factor and $G_{\rm P}$ for $7\,{\rm ML}$ MgO, the $G_{\rm P}$ for $13\,{\rm ML}$ MgO is approximately estimated as $G_{\rm P}(7\,{\rm ML}\,\,{\rm MgO}) \times \exp(-2\kappa d) \approx 1.17 \times 10^{-6}\,e^2/h$, which is close to $2.64 \times 10^{-6}\,e^2/h$ (Table \ref{tab2}) obtained in the actual transport calculation. Note that this type of comparison makes sense for the order of the conductance in the present case. This is because the present TMR comes from the interfacial resonance effect, in which many ${\bf k}_\parallel$ points on the rings around $\Gamma$ [Fig. \ref{CoNiMgO}(c)] provide a large contribution to $G_{\rm P}$ and each ${\bf k}_\parallel$ point has a different value of $\kappa$.

We also studied the interfacial LDOSs and ${\bf k}_\parallel$-resolved conductance of CoPt/MgO/CoPt(111) with the second highest TMR ratio [Figs. \ref{CoPtMgO}(a) to \ref{CoPtMgO}(e)]. In this case, the interfacial antibonding related to the high TMR ratio is formed in the minority-spin state, not the majority-spin state. As shown in Fig. \ref{CoPtMgO}(b), O $p_x$ and $p_y$ minority-spin states have large LDOSs at the Fermi level owing to the antibonding with Co $d_{zx}$ and $d_{yz}$ states. These interfacial states provide a high TMR ratio through the interface resonant tunneling. Actually, the conductance with the largest contribution to the high TMR ratio is that in the minority-spin state $G_{{\rm P},\downarrow}({\bf k}_\parallel)$ [Fig. \ref{CoPtMgO}(c)], whose ${\bf k}_\parallel$ dependence can be reproduced by that of the LDOSs in the interfacial O $p_x$ and $p_y$ minority-spin states [Figs. \ref{CoPtMgO}(d) and \ref{CoPtMgO}(e)].

Such a difference in the spin channel contributing to the high TMR ratio between the CoNi- and CoPt-based MTJs comes from different exchange splittings in the interfacial Co $d_{zx}$ and $d_{yz}$ states. By comparing Figs. \ref{CoNiMgO}(a) and \ref{CoPtMgO}(a), we can easily see that the exchange splitting in the $d_{zx}$ and $d_{yz}$ states in the CoNi-based MTJ is clearly smaller than that in the CoPt-based MTJ. In fact, the magnetic moment projected onto each $d$ orbital in the interfacial Co atom was estimated in both MTJs. We obtained 0.96\,$\mu_{\rm B}$ in the $d_{zx}$ and $d_{yz}$ orbitals for the CoNi-based MTJ and 1.10\,$\mu_{\rm B}$ for the CoPt-based MTJ. In the other $d$ orbitals, the difference in the projected magnetic moment was found to be quite small. Therefore, in the CoNi-based MTJ, the $d_{zx}$ and $d_{yz}$ majority-spin states have finite majority-spin LDOSs at the Fermi level, leading to the large O $p_x$ and $p_y$ majority-spin LDOSs through the interfacial antibonding [Fig. \ref{CoNiMgO}(b)]. In contrast, the CoPt-based MTJ has negligibly small $d_{zx}$ and $d_{yz}$ majority-spin LDOSs at the Fermi level owing to the larger exchange splitting [Fig. \ref{CoPtMgO}(a)], which provides the dominance of the minority-spin LDOSs in the interfacial O $p$ states [Fig. \ref{CoPtMgO}(b)].

Although not shown here, we confirmed that the high TMR ratio in the CoPd-based MTJ (2172\%) can also be explained by the interface resonant tunneling of the interfacial O $p_x$ and $p_y$ minority-spin states. Our present study revealed that not only the Co/MgO/Co(111) MTJ \cite{2020Masuda-PRB} but also several (111)-oriented MTJs with Co-based $L1_1$ alloys exhibit high TMR ratios due to the interface resonant tunneling. This fact allows us to expect that such a mechanism may be universal for high TMR ratios in (111)-oriented MTJs.

\begin{figure}
\includegraphics[width=8.9cm]{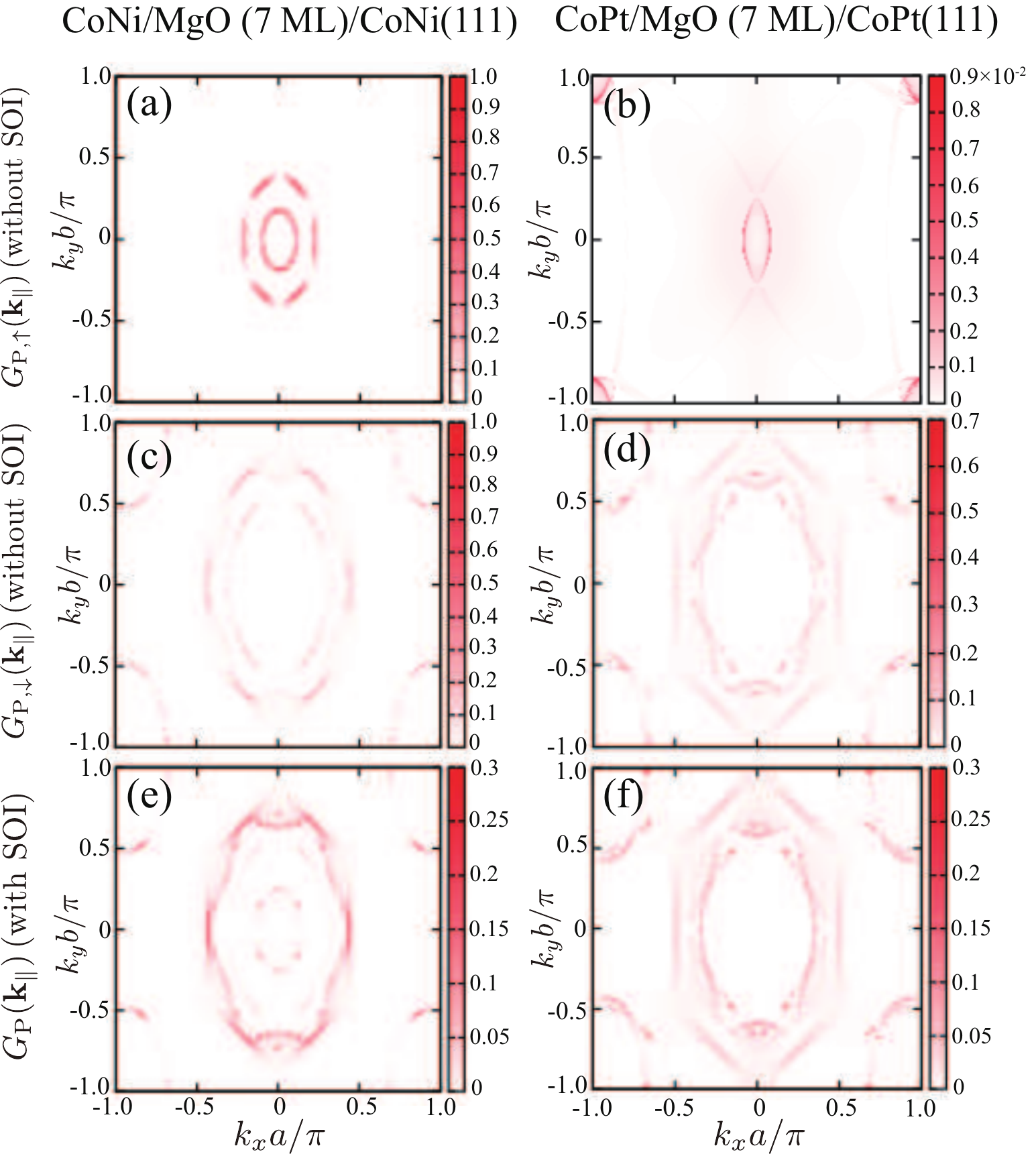}
\caption{\label{conductance_SOI} The comparison of ${\bf k}_\parallel$-resolved conductances between the cases with and without SOI in ${\rm CoNi}/{\rm MgO}\,(7\,{\rm ML})/{\rm CoNi}(111)$ (a),\,(c),\,(e) and ${\rm CoPt}/{\rm MgO}\,(7\,{\rm ML})/{\rm CoPt}(111)$ (b),\,(d),\,(f) with the parallel configuration of magnetization. (a,\,b) Majority-spin conductances $G_{{\rm P},\uparrow}({\bf k}_\parallel)$ and (c,\,d) Minority-spin conductances $G_{{\rm P},\downarrow}({\bf k}_\parallel)$ for the case without SOI. (e,\,f) Conductances $G_{\rm P}({\bf k}_\parallel)$ for the case with SOI. The unit of the color bar is $e^2/h$ in all panels.}
\end{figure}
\subsection{Effect of the SOI on TMR ratios}
To discuss the effect of the SOI on the TMR effect, we tried the calculation of conductances considering the SOI in CoNi/MgO/CoNi(111) and CoPt/MgO/CoPt(111), which were shown to have quite high TMR ratios for the case without the SOI. Figures \ref{conductance_SOI}(e) and \ref{conductance_SOI}(f) show the ${\bf k}_\parallel$-resolved conductances for the parallel configuration of magnetization in CoNi/MgO/CoNi(111) and CoPt/MgO/CoPt(111), respectively. We see that the feature of the ${\bf k}_\parallel$ dependence can be naturally understood by combining those of the majority- and minority-spin conductances in the absence of the SOI [Figs. \ref{conductance_SOI}(a) to \ref{conductance_SOI}(d)]. In addition, the difference in the ${\bf k}_\parallel$-averaged conductance between the cases with and without the SOI is not so large; for example, we obtained $G_{\rm P}=4.98\times10^{-3}\, e^2/h$ in CoPt/MgO/CoPt(111) with the SOI, which is $25\%$ smaller than $G_{\rm P}=G_{{\rm P},\uparrow}+G_{{\rm P},\downarrow}=6.61\times10^{-3}\, e^2/h$ in the case without the SOI. On the other hand, in the antiparallel configuration of magnetization, the self-consistent-field calculation for the supercell did not converge in both systems within a realistic calculation time. Thus, we could not estimate TMR ratios directly from such calculations.

However, we can approximately estimate the effect of the SOI on TMR ratios only from the obtained parallel conductance $G_{\rm P}$. We here consider the case of CoPt/MgO/CoPt(111), where $G_{\rm P}$ is decreased to $75\%$ of the original value by including the SOI as shown above. Let us here introduce an assumption that the conductance is proportional to the product of LDOSs at the left and right interfaces. This is the similar assumption used in the derivation of the well-known Julliere formula \cite{1975Julliere-PL}; however, the bulk DOSs of electrodes in the original assumption are replaced by the interfacial LDOSs in the present case, since the interfacial resonant tunneling is found to be the origin of the present TMR. Based on this picture, the parallel conductance is given by $G_{\rm P} \propto D_{{\rm L}\uparrow}D_{{\rm R}\uparrow}+D_{{\rm L}\downarrow}D_{{\rm R}\downarrow} \approx D_{{\rm L}\downarrow}D_{{\rm R}\downarrow}$, where $D_{{\rm L}\uparrow(\downarrow)}$ and $D_{{\rm R}\uparrow(\downarrow)}$ are the majority-spin (minority-spin) LDOSs at $E_{\rm F}$ in the left and right interfaces, respectively. Here, we assumed that the minority-spin LDOSs $D_{{\rm L(R)}\downarrow}$ are sufficiently large compared to the majority-spin ones $D_{{\rm L(R)}\uparrow}$ as seen from Figs. \ref{CoPtMgO}(a) and \ref{CoPtMgO}(b) \cite{note1_dos}. From $G_{\rm P}\approx D_{{\rm L}\downarrow}D_{{\rm R}\downarrow}$ and the fact that the SOI decreases $G_{\rm P}$ to $75\%$ of the original value, it is estimated that each of $D_{{\rm L}\downarrow}$ and $D_{{\rm R}\downarrow}$ becomes $\sqrt{0.75}$ times smaller than the original value. Since the antiparallel conductance is given by $G_{\rm AP} \propto D_{{\rm L}\uparrow}D_{{\rm R}\downarrow}+D_{{\rm L}\downarrow}D_{{\rm R}\uparrow}$, ${\rm TMR\, ratio}\, (\%) \approx 100 \times (G_{\rm P}/G_{\rm AP})$ becomes $0.75/\sqrt{0.75}=\sqrt{0.75}$ times smaller than the original value \cite{note2_dos}. Therefore, by using the original TMR ratio in Table \ref{tab1}, it is concluded that the TMR ratio of CoPt/MgO/CoPt(111) decreases from $2534\%$ to $2194\%$ by including the SOI. Although this is a rough estimation, we can expect that the SOI does not affect the TMR ratio significantly. Note here that such a small effect of the SOI would be related to the fact that the interface is made by Co and O atoms (not containing Pt atoms with a large SOI). As shown in Sec. \ref{modelmethod} A, such a Co-O interface is energetically stable from the theoretical point of view. However, another interface such as Pt-O may occur in actual experiments, where the effect of the SOI might be large. Thus, the systematic analysis of TMR ratios fully including the SOI should be addressed in future studies.

\subsection{Large PMA and its correlation with perturbation processes}
\begin{figure}
\includegraphics[width=7.8cm]{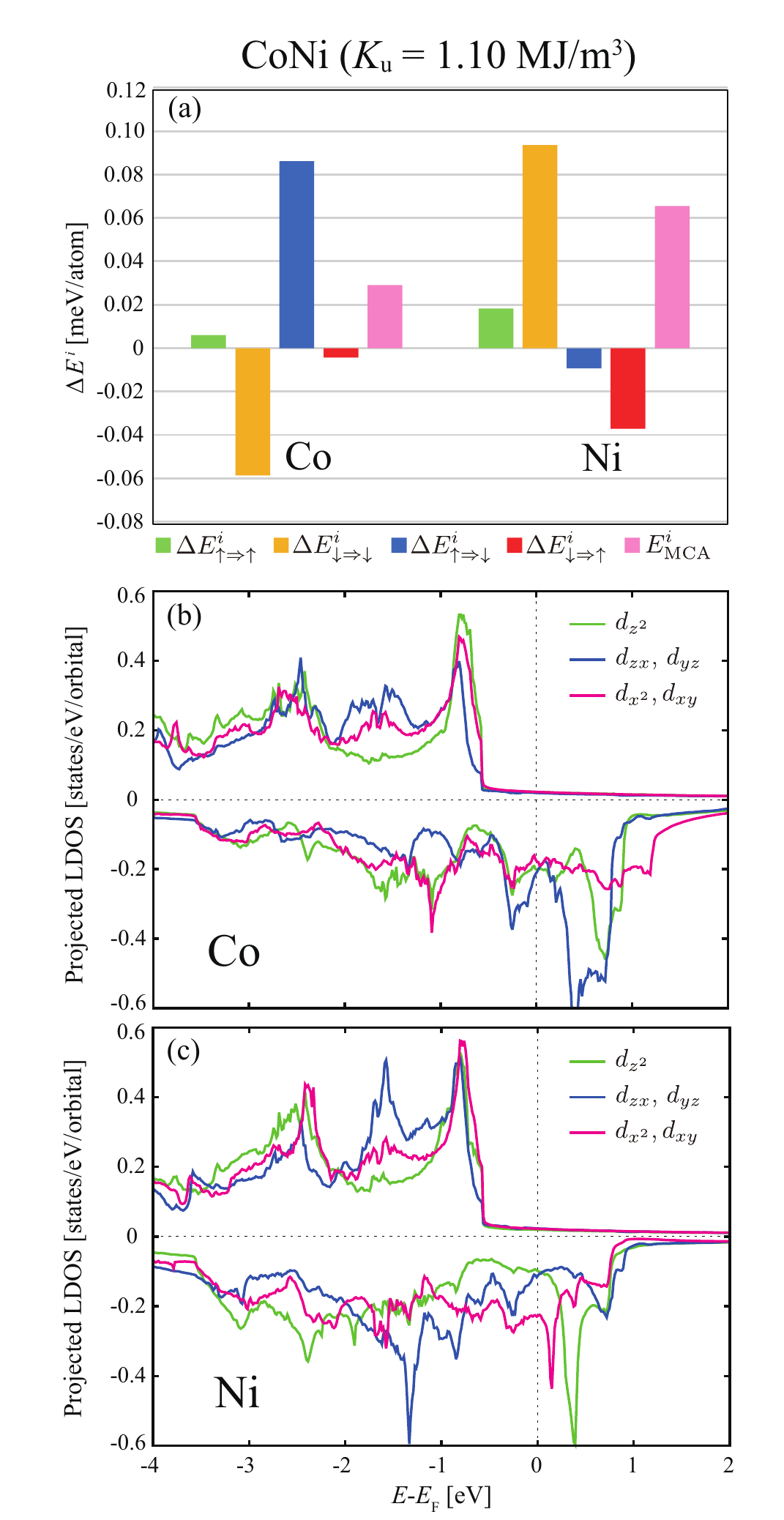}
\caption{\label{CoNi_pert_LDOS} (a) Results of second-order perturbation analysis on the PMA in $L1_1$ CoNi. (b,\,c) Projected LDOS for Co and Ni atoms in $L1_1$ CoNi, where $d_{3z^2-r^2}$ and $d_{x^2-y^2}$ are abbreviated as $d_{z^2}$ and $d_{x^2}$, respectively.}
\end{figure}
\begin{figure}
\includegraphics[width=7.8cm]{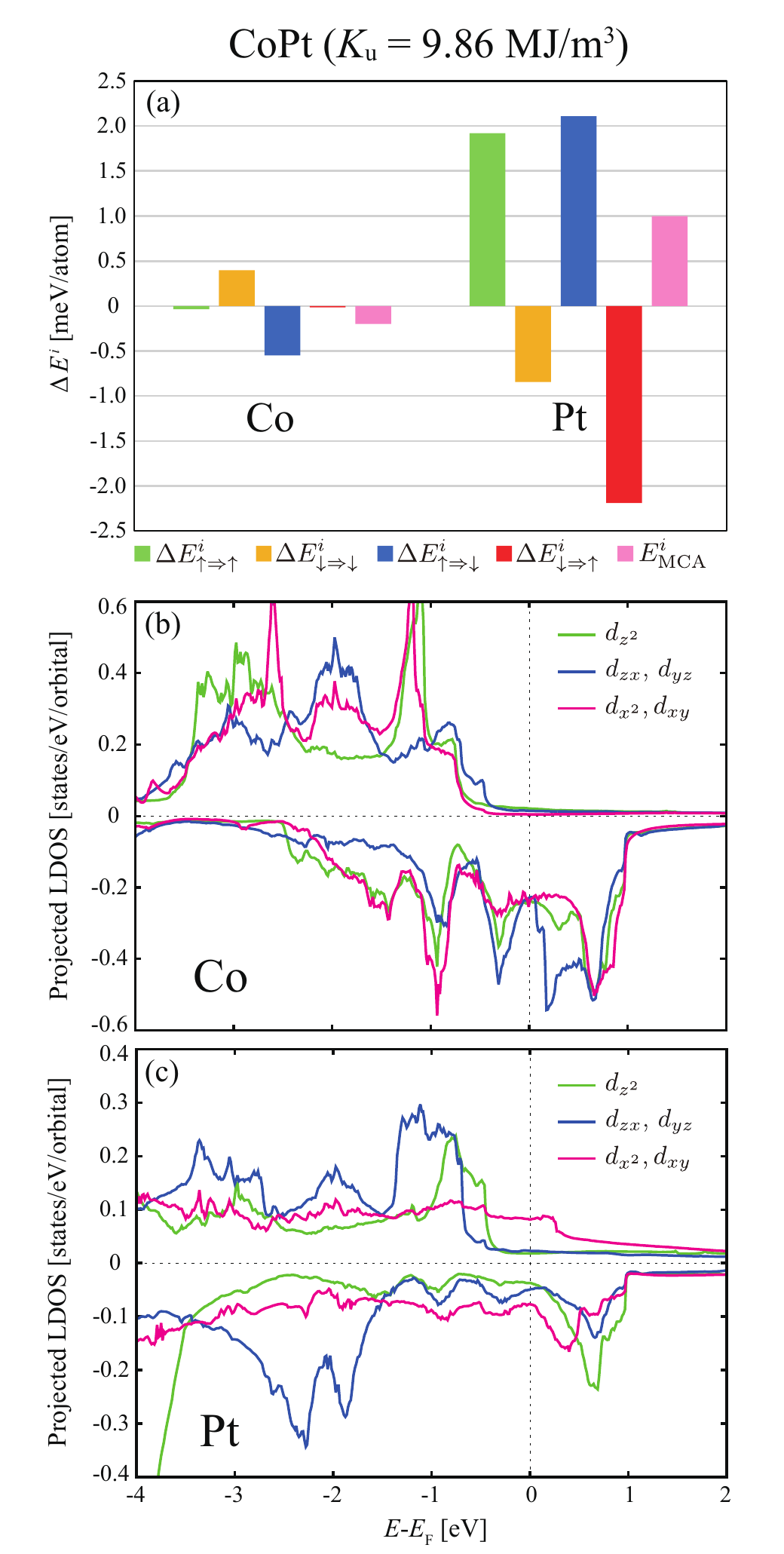}
\caption{\label{CoPt_pert_LDOS} The same as Fig. \ref{CoNi_pert_LDOS}, but for $L1_1$ CoPt.}
\end{figure}
We listed the obtained values of $K_{\rm u}$ in Table \ref{tab1}. All the alloys except NiPt have positive $K_{\rm u}$ indicating a tendency toward PMA. Among them, CoPt possesses the largest value close to 10\,MJ/m$^3$. In this section, we discuss the origin of $K_{\rm u}$ in CoNi and CoPt as representatives based on the second-order perturbation analysis of the magnetocrystalline anisotropy. Here, we used $\xi_{\rm Co}=69.4\,{\rm meV}$, $\xi_{\rm Ni}=87.2\,{\rm meV}$, and $\xi_{\rm Pt}=523.8\,{\rm meV}$ as the coupling constants of the SOI $\xi_i$. We also set the Wigner-Seitz radius of each atom to $r_{\rm Co}=1.302$\,\AA, $r_{\rm Ni}=1.286$\,\AA, and $r_{\rm Pt}=1.455$\,\AA\, for obtaining projected wave functions used in the calculation. All these values are those in the pseudopotential files in the {\scriptsize VASP} code.

Figure \ref{CoNi_pert_LDOS}(a) shows the results of the second-order perturbation analysis of $K_{\rm u}$ in CoNi. We see that Ni has a much larger positive $E^i_{\rm MCA}$ than Co and contributes dominantly to the PMA. In the Co atom, $\Delta E^i_{\downarrow \Rightarrow \downarrow}$ and $\Delta E^i_{\uparrow \Rightarrow \downarrow}$ have large values but with opposite signs, leading to a small $\Delta E^i_{\rm MCA}$. In contrast, in the Ni atom, the spin-conserving term $\Delta E^i_{\downarrow \Rightarrow \downarrow}$ in the minority-spin channel is positive and much larger than the other terms, giving a large positive $E^i_{\rm MCA}$. This is consistent with the LDOSs of Ni shown in Fig. \ref{CoNi_pert_LDOS}(c), where the minority-spin state has large values around $E_{\rm F}$, while the majority-spin state has only small values. It is known that the expression of $\Delta E^i_{\downarrow \Rightarrow \downarrow}$ within the second-order perturbation theory is analytically given by
\begin{equation}
\Delta E^{i}_{\downarrow \Rightarrow \downarrow} = \xi_{i}^2 \sum_{u_{\downarrow},o_{\downarrow}} \frac{|\langle u_{\downarrow} |L^{i}_{z}| o_{\downarrow} \rangle |^{2} - |\langle u_{\downarrow} |L^{i}_{x}| o_{\downarrow} \rangle |^{2}}{\epsilon_{u_{\downarrow}}-\epsilon_{o_{\downarrow}}}, \label{eq:dndn}
\end{equation}
where $L^{i}_{\alpha}\,(\alpha=x,z)$ is the local angular momentum operator at an atomic site $i$, and $| o_{\sigma} \rangle$ ($| u_{\sigma} \rangle$) is a local occupied (unoccupied) state with spin $\sigma$ and energy $\epsilon_{o_{\sigma}}$ ($\epsilon_{u_{\sigma}}$) \cite{1993Wang-PRB}. This expression indicates that the matrix element of $L^{i}_{z}$ gives a positive contribution to $\Delta E^{i}_{\downarrow \Rightarrow \downarrow}$ while that of $L^{i}_{x}$ gives a negative contribution. Actually, we confirmed that $\langle d_{x^2-y^2},\downarrow |L^{i}_{z}| d_{xy},\downarrow \rangle$ and $\langle d_{xy},\downarrow |L^{i}_{z}| d_{x^2-y^2},\downarrow \rangle$ have large values in our perturbation calculation, which is consistent with large minority-spin LDOSs in the $d_{x^2-y^2}$ and $d_{xy}$ states shown in Fig. \ref{CoNi_pert_LDOS}(c).

Figure \ref{CoPt_pert_LDOS} presents the results for CoPt. From the perturbation analysis [Fig. \ref{CoPt_pert_LDOS}(a)], we find that in all spin-transition processes Pt has much larger anisotropy energy than Co, meaning that the PMA in CoPt mainly comes from the anisotropy in Pt. In the Pt atom, a large positive anisotropy $\Delta E^i_{\uparrow \Rightarrow \downarrow}$ is found in the $\uparrow \Rightarrow \downarrow$ spin-flip process, but this is canceled out by $\Delta E^i_{\downarrow \Rightarrow \uparrow}$ in the other spin-flip process. Thus, the dominant contribution to the large positive anisotropy in Pt is given by $\Delta E^i_{\uparrow \Rightarrow \uparrow}$ in the $\uparrow \Rightarrow \uparrow$ spin-conserving process. Similar to Eq. (\ref{eq:dndn}), the analytical expression of $\Delta E^i_{\uparrow \Rightarrow \uparrow}$ is given as follows \cite{1993Wang-PRB}:
\begin{equation}
\Delta E^{i}_{\uparrow \Rightarrow \uparrow} = \xi_{i}^2 \sum_{u_{\uparrow},o_{\uparrow}} \frac{|\langle u_{\uparrow} |L^{i}_{z}| o_{\uparrow} \rangle |^{2} - |\langle u_{\uparrow} |L^{i}_{x}| o_{\uparrow} \rangle |^{2}}{\epsilon_{u_{\uparrow}}-\epsilon_{o_{\uparrow}}}, \label{eq:upup}
\end{equation}
from which the matrix element of $L_z$ is found to give a positive contribution to $\Delta E^{i}_{\uparrow \Rightarrow \uparrow}$. As clearly seen in Fig. \ref{CoPt_pert_LDOS}(c), the $d_{x^2-y^2}$ and $d_{xy}$ states have much larger LDOSs than the other $d$ states around $E_{\rm F}$ in the majority-spin channel. Such LDOSs yield large values of $\langle d_{x^2-y^2},\uparrow |L^{i}_{z}| d_{xy},\uparrow \rangle$ and $\langle d_{xy},\uparrow |L^{i}_{z}| d_{x^2-y^2},\uparrow \rangle$, leading to a large positive $\Delta E^i_{\uparrow \Rightarrow \uparrow}$. The importance of the $\uparrow \Rightarrow \uparrow$ term is also found in Pt of $L1_0$ FePt with large PMA \cite{2016Ueda-APL} and is a feature in ordered alloys with Pt atoms.

Conventionally, PMA has been explained with the help of the Bruno theory \cite{1989Bruno-PRB}, which states that PMA mainly comes from the anisotropy of the orbital magnetic moment, namely, the spin-conserving term $\Delta E^{i}_{\downarrow \Rightarrow \downarrow}$ in Eq. (\ref{eq:E2MCA_decomp}). This theory is applicable to typical ferromagnets with large exchange splittings, since such ferromagnets have almost occupied majority-spin states, and only minority-spin states are located close to the Fermi level. In contrast, many recent studies on PMA \cite{2016Ueda-APL,2017Miwa-NatCom,2018Masuda-PRB,2019Okabayashi-APL,2017Seki-JPSJ,2018Okabayashi-SciRep,2020Okabayashi-SciRep} focused on its unconventional mechanism due to the spin-flip terms $\Delta E^{i}_{\uparrow \Rightarrow \downarrow}$ and $\Delta E^{i}_{\downarrow \Rightarrow \uparrow}$ in Eq. (\ref{eq:E2MCA_decomp}). These terms can be interpreted in terms of the quadrupole moment and provide novel physical insight into PMA. Up to now, it has been shown that the spin-flip terms play a significant role for PMA in various systems including ferromagnet/MgO interfaces and ferromagnetic multilayers \cite{2016Ueda-APL,2017Miwa-NatCom,2018Masuda-PRB,2019Okabayashi-APL,2017Seki-JPSJ,2018Okabayashi-SciRep,2020Okabayashi-SciRep}. In the present study, we obtained large values of spin-flip terms in $L1_1$ CoNi and CoPt. However, as mentioned above, $\Delta E^{i}_{\uparrow \Rightarrow \downarrow}$ is canceled by $\Delta E^{i}_{\downarrow \Rightarrow \downarrow}$ in CoNi and two types of spin-flip terms are canceled with each other in CoPt. Therefore, the unconventional physical picture is not suitable to explain PMA in the present CoNi and CoPt. A similar cancellation of the spin-flip terms has also been reported recently in an FeIr/MgO system \cite{2019Miwa-PRB}.

\section{summary}
We theoretically investigated the TMR effect and magnetocrystalline anisotropy in (111)-oriented MTJs with $L1_1$ alloys based on the first-principles calculations. Our transport calculation showed that the MTJs with Co-based alloys (CoNi, CoPt, and CoPd) have high TMR ratios over 2000\%, which are attributed to the interface resonant tunneling. We also found that the tunneling mainly occurs in the majority-spin channel in the CoNi-based MTJ while it occurs in the minority-spin channel in the CoPt-based MTJ, meaning that different spin channels provide dominant contributions to the high TMR ratios in different systems. This can be understood from the different exchange splittings in the $d_{zx}$ and $d_{yz}$ states of interfacial Co atoms contributing to the TMR effect through antibonding with O $p_x$ and $p_y$ states. The analysis of the magnetocrystalline anisotropy revealed that many $L1_1$ alloys have large PMA and CoPt has the largest value of $K_{\rm u} \approx 10\,{\rm MJ/m^3}$. Through a detailed second-order perturbation calculation, we clarified that the large PMA in CoPt and CoNi is attributed to the spin-conserving perturbation processes around the Fermi level. All these findings would be useful for understanding experimental results in (111)-oriented MTJs, which will be obtained in future studies.

\begin{acknowledgments}
The authors are grateful to S. Takahashi and K. Nawa for helpful discussions and critical comments. This work was partly supported by Samsung Electronics, Grant-in-Aids for Scientific Research (S) (Grant No. JP16H06332 and No. JP17H06152), Scientific Research (B) (Grant No. JP20H02190), and for Early-Career Scientists (Grant No. JP20K14782) from the Ministry of Education, Culture, Sports, Science and Technology, Japan, and NIMS MI$^2$I. The crystal structures were visualized using {\scriptsize VESTA} \cite{2011Momma_JAC}.
\end{acknowledgments}


\begin{thebibliography}{99}
\bibitem{note1_dos} Note that in addition to $|D_{{\rm L(R)},\uparrow}| \ll |D_{{\rm L(R)},\downarrow}|$, the effect of the SOI on $D_{{\rm L(R)},\uparrow}$ is considered to be much smaller than that on $D_{{\rm L(R)},\downarrow}$, since the energy dependence of $D_{{\rm L(R)},\uparrow}$ around $E_{\rm F}$ is much flatter than that of $D_{{\rm L(R)},\downarrow}$ as seen in Figs. \ref{CoPtMgO}(a) and \ref{CoPtMgO}(b).
\bibitem{note2_dos} We approximately neglected the change in $D_{{\rm L(R)},\uparrow}$ due to the SOI using the fact mentioned in Ref. \cite{note1_dos}.
\end{thebibliography}

\end{document}